\documentclass[useAMS,usenatbib]{mn2e}
\usepackage{graphicx}

\title[High resolution infrared spectra of NGC~6440 and NGC~6441]
{High resolution infrared spectra of NGC~6440 and NGC~6441:
\thanks{Data presented herein were obtained at the W.M.Keck Observatory,
        which is operated as a scientific partnership among the California
        Institute of Technology, the University of California, and the National
        Aeronautics and Space Administration. The Observatory was made possible
        by the generous financial support of the W.M. Keck Foundation.}
two massive Bulge Globular Clusters}

\author[Origlia, Valenti \& Rich]{L. Origlia$^1$, E. Valenti$^{2,1,3}$,
        R.~M. Rich$^3$  \\
 $^1$ INAF-Osservatorio Astronomico di Bologna, Via Ranzani 1, I-40127 Bologna,
      Italy, e-mail livia.origlia@oabo.inaf.it \\
 $^2$ ESO - European Southern Observatory, Vitacura, Santiago, Casilla 19001, Chile,
      e-mail evalenti@eso.org\\  
      $^3$ Department of Physics and Astronomy, University of California
      at Los Angeles, Los Angeles, CA 90095-1562, e-mail rmr@astro.ucla.edu \\
       }
\date{\today}

\begin{document}
\pagerange{\pageref{firstpage}--\pageref{lastpage}} \pubyear{2008}
\maketitle
\label{firstpage}

\begin{abstract}
Using the NIRSPEC spectrograph at Keck II, we have obtained
infrared echelle spectra covering the $1.5-1.8~\mu \rm m$ range for
giant stars in the massive bulge globular clusters NGC~6440 and
NGC~6441. We report the first high dispersion abundance for NGC~6440, [Fe/H]=$-0.56\pm0.02$ and 
we find [Fe/H]=$-0.50\pm0.02$ for the blue HB cluster NGC~6441.
We measure an average $\alpha$-enhancement of $\approx+0.3$~dex
in both clusters,
consistent with previous measurements of other metal rich bulge clusters, 
and favoring the scenario of a rapid bulge formation and chemical enrichment.
We also measure very low  $\rm ^{12}C/^{13}C$ isotopic ratios 
($\approx 5\pm 1$),
suggesting that extra-mixing mechanisms 
are at work during evolution along the Red Giant Branch 
also in the high metallicity regime.
We also measure 
Al abundances, finding 
average $\rm [Al/Fe]=0.45\pm0.02$ and
$\rm [Al/Fe]=0.52\pm0.02$ in NGC~6440 and NGC~6441, respectively, 
and some Mg-Al anti-correlation in NGC~6441. 
We also measure radial velocities v$_r$=--76$\pm$3 km/s and v$_r$=+14$\pm$3 km/s 
and velocity dispersions $\sigma$=9$\pm$2 km/s and $\sigma$=10$\pm$2 km/s, 
in NGC~6440 and NGC~6441, respectively.

\end{abstract}

%______________________________________________________________________
\begin{figure*}
\centering
 \leavevmode
 \includegraphics[width=8.5cm]{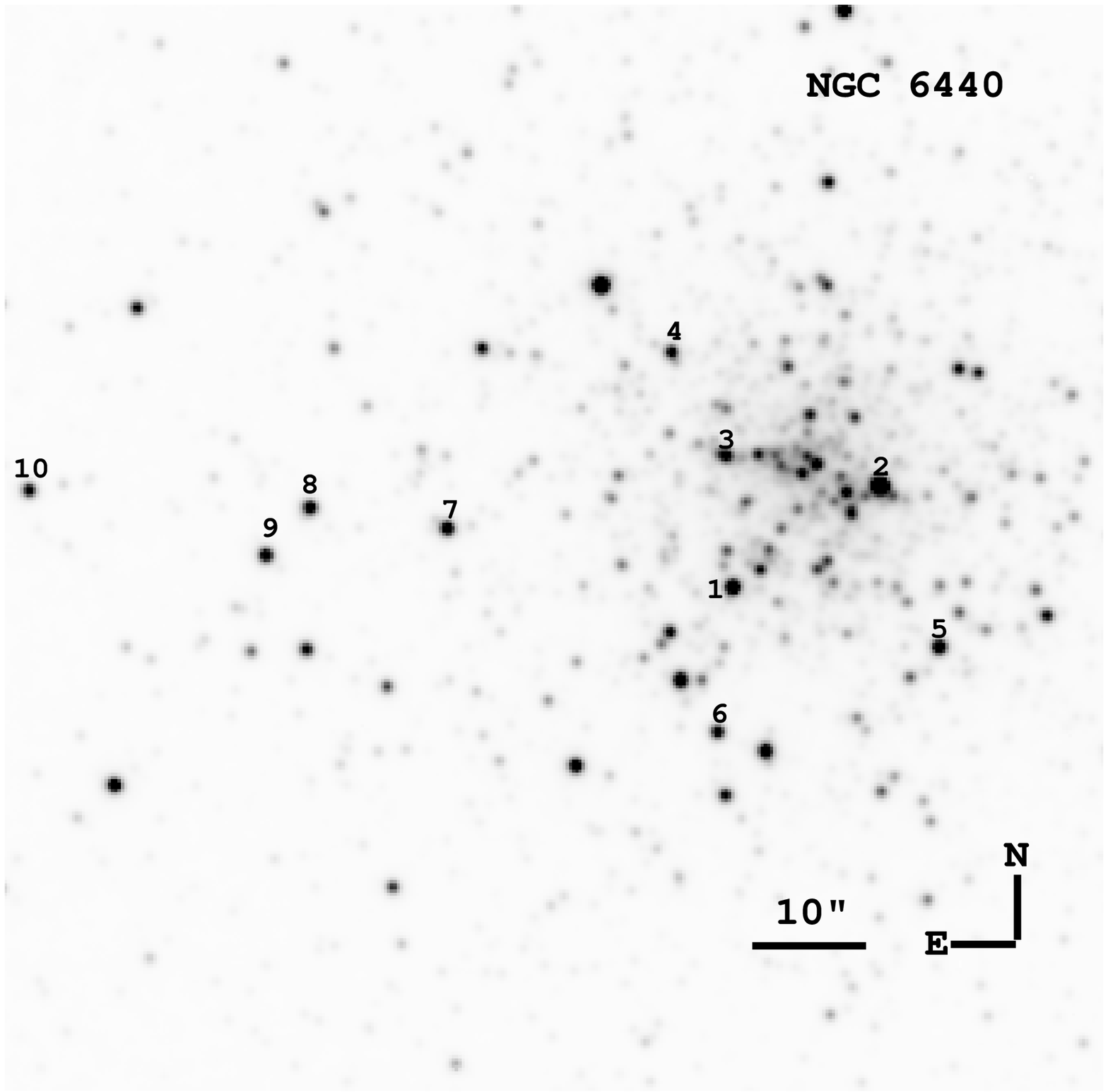}
 \hfil
 \includegraphics[width=8.5cm]{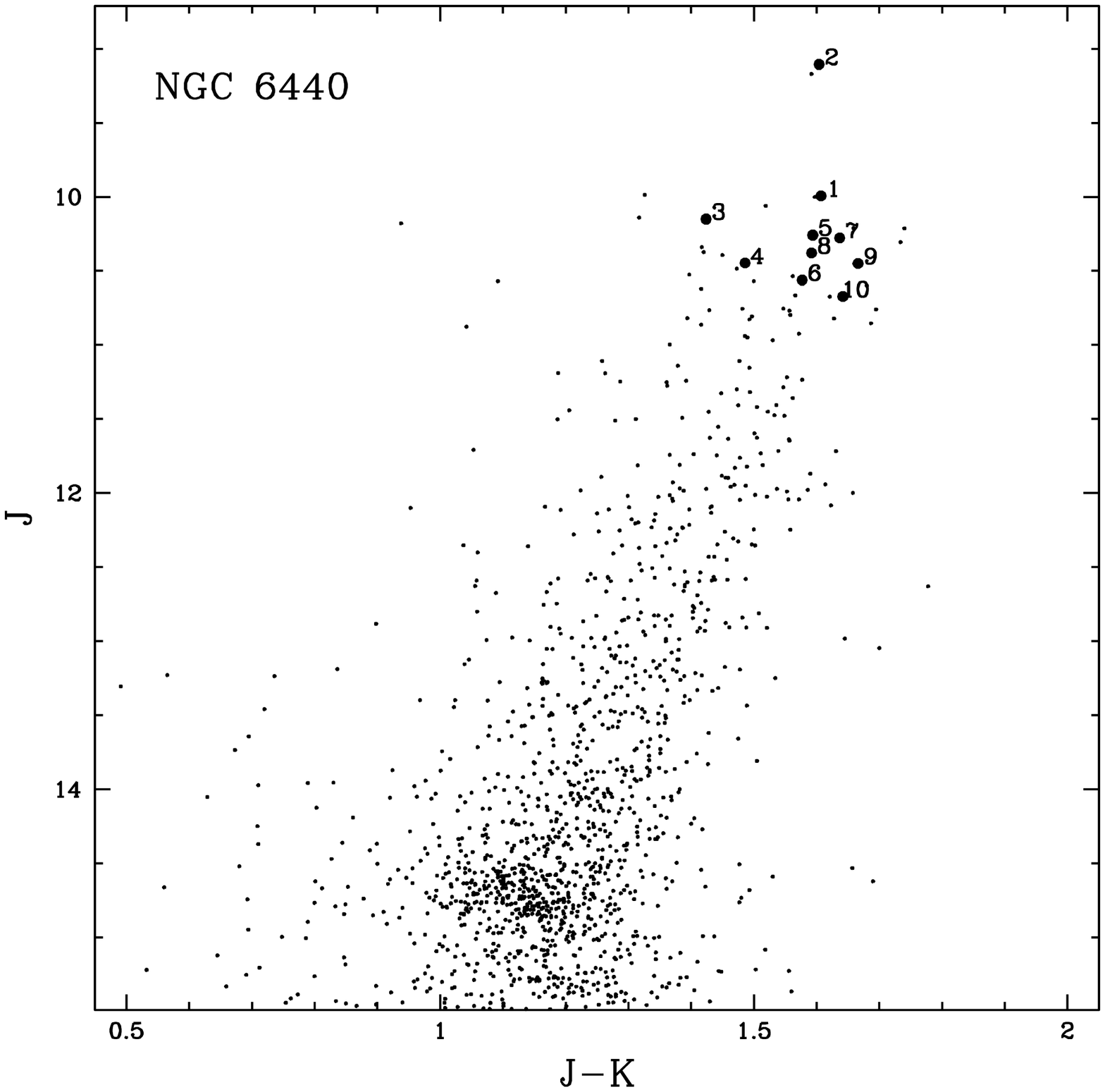}
\caption{
J band image of the core region (left panel) and the J,(J-K) 
color-magnitude diagram (right panel) of NGC~6440 \citep{val04}. 
The stars spectroscopically observed are numbered (cf. Table~1).
}
\label{fig1}
\end{figure*}
%________________________________________________________________

\section{Introduction}

Over the past few years we have been undertaking
a high-resolution spectroscopic survey of the Galactic 
bulge in the near-IR using NIRSPEC, a high throughput infrared (IR) echelle
spectrograph at the Keck Observatory \citep{ml98}.
The near IR spectral range is well suited to measure the obscured stellar populations 
in the inner Galaxy.
H--band (1.5--1.8 $\mu$m) spectra
of bright giants in the bulge globular clusters and field population are ideal
for detailed abundance analysis of Fe, 
C, O and other $\alpha $-elements, using the approach of synthesizing the
entire spectrum.  
The abundance distributions in the {\it cluster} and {\it field}
populations are important in constraining the history and timescale of bulge formation and 
chemical enrichment \citep{rich90,mw97,mrm99}.

We have used this method to derive abundances for
eight bulge globular clusters in the inner and outer bulge: 
the resulting abundances for NGC~6553 
and Liller~1 are given in \citet{ori02}, for 
Terzan~4 and Terzan~5 in \citet{ori04}, for NGC~6342 and NGC6528 in \citet{ori05a}, 
and for NGC~6539 and UKS~1 in \citet{ori05b}.
We also measured detailed abundances of bulge giants in the Baade's window \citep{rich05} and 
in the inner field at $(l,b)=(0,-1)$ \citep{rich07}.

We find $\alpha$-enhancement at a level of a factor between 2 and 3 over 
the whole range of metallicity spanned by the bulge clusters in our survey, 
from [Fe/H]$\approx$--1.6 (cf. Terzan~4) up to [Fe/H]$\approx$--0.2 (cf. Terzan~
5).  Such an enhancement has been also found in the observed bulge fields, 
without any evidence of vertical abundance and abundance pattern gradients.

In this paper we present the high resolution
IR spectra and the abundance analysis of  
bright giants in NGC~6440 and NGC~6441,
two massive bulge globular clusters 
of the inner bulge, located at $(l,b)=(7.7,3.8)$ and $(l,b)=(353.5,-5.0)$, respectively 
\citep{har96}.

During the 1990s, NGC~6440 has been photometrically observed in the optical \citep{ort94}
and near IR \citep{kuc95,min95}, suggesting an overall [Fe/H] between 
1/3 and half solar and E(B-V)$\approx1.0\pm0.1$.
Low resolution spectroscopy \citep{az88,min95,ori97,fro01} provides very similar metallicities. 
More recently, \citet{val04} presented new near IR photometry, 
giving very similar [Fe/H]=-0.49 and slightly higher E(B-V)=1.15, 
but in very good agreement with \citet{sch98}.
No high spectral resolution measurements of this cluster exists so far.

NGC~6441 can be regarded as a twin cluster to NGC~6440, being also very massive and 
with an overall metallicity between 1/3 and half solar
\citep{az88,hr99,val04}, but with an anomalous
blue horizontal branch \citep{rich97} and RR Lyrae population 
\citep[][ and references therein]{cle05} for its high metallicity.
Given its lower 
E(B-V)$\approx0.5\pm0.1$ extinction \citep[see e.g.][ for a review]{gra06} 
compared to NGC~6440,
this cluster has been also studied at high spectral resolution in the optical 
\citep{gra06,gra07}, finding [Fe/H]=-0.39, [$\alpha$/Fe] enhancement and Na-O anti-correlation.

Our observations and data reduction follow in Sect.~2.
Sect.~3 presents our abundance analysis 
Sect.~4 the resulting chemical abundances and radial velocities.
We discuss our findings in Sect.~5.

%______________________________________________________________________
\begin{figure*}
\centering
 \leavevmode
 \includegraphics[width=8.5cm]{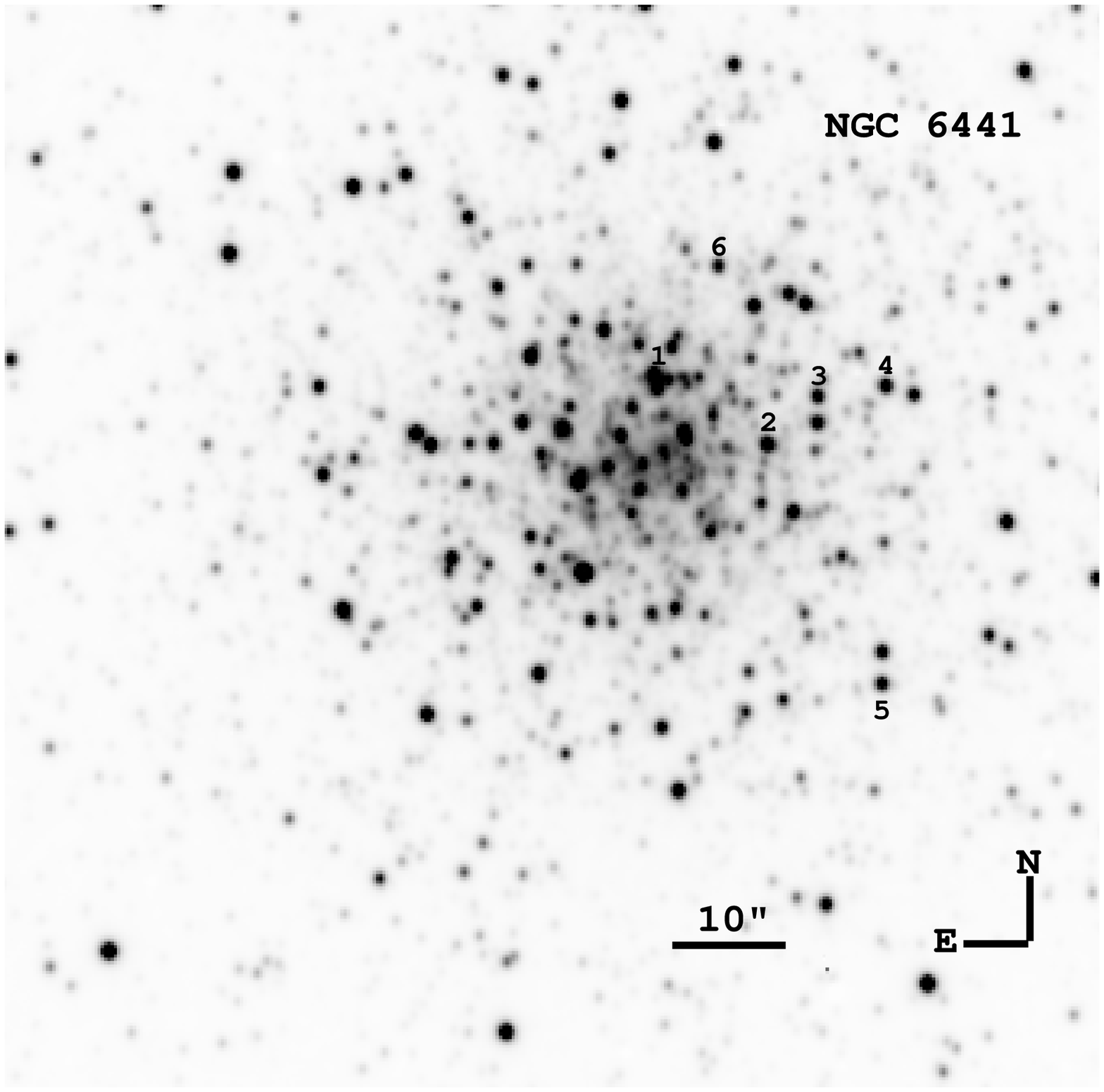}
 \hfil
 \includegraphics[width=8.5cm]{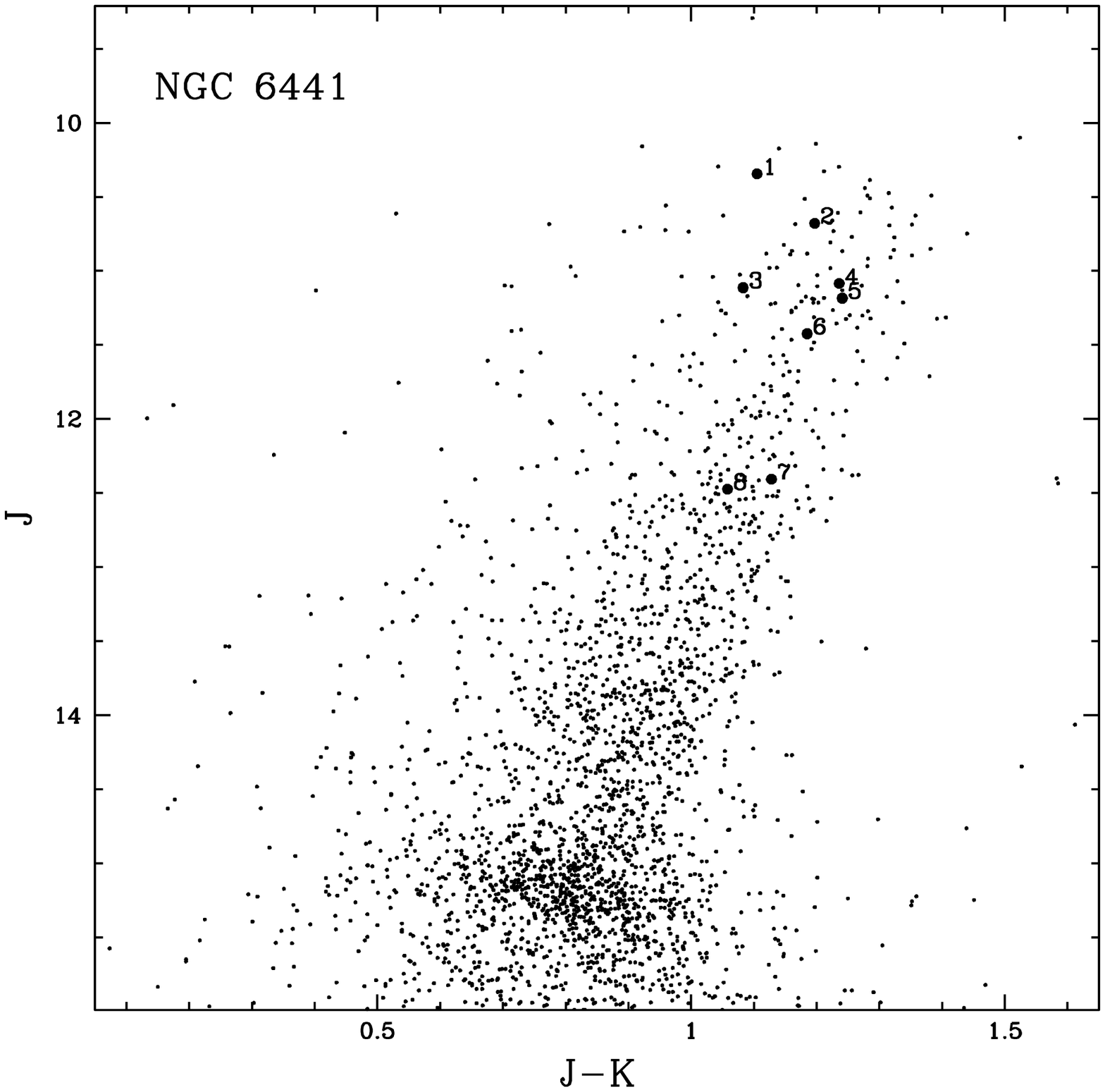}
\caption{
H band image of the core region (left panel) and the J,(J-K) 
color-magnitude diagram (right panel) of NGC~6441 \citep{val04}. 
The stars spectroscopically observed are numbered (cf. Table~2).
}
\label{fig2}
\end{figure*}
%________________________________________________________________

\section{Observations and Data Reduction}

Near infrared, high-resolution echelle spectra of bright giants
in the core of the bulge globular
clusters NGC~6440 and NGC~6441
have been acquired
with the infrared spectrograph NIRSPEC \citep{ml98}
mounted at the Nasmyth focus of the Keck~II telescope.
The high resolution echelle mode, with a slit width of $0\farcs43$
(3 pixels) and a length of 12-24\arcsec\
and the standard NIRSPEC-5 setting, which
covers most of the 1.5--1.8 micron H-band,
have been selected.  Typical exposure times (on source)
are $\approx$8 minutes.
NGC~6440 was observed on July 2003 and April 2007, while NGC~6441 on July and 2002 and May 2006.
A total of 10 and 8 giants have been measured, respectively.
Figs.~\ref{fig1},~\ref{fig2} show the IR image and the color-magnitude diagram \citep{val04} of 
the central region of NGC~6440 and NGC~6441, respectively, with marked the giant stars spectroscopically 
observed.
The selected stars are within a magnitude from the RGB Tip, 
that is sufficiently bright to give a good signal to noise in a relatively short integration time 
and suffucently below the nominal RGB tip in order to minimize possible 
contamination by brighter AGB stars.
Moreover, the selection of stars in the core region maximizes the membership probability, although 
a spectroscopic confirmation is always required.
In NGC~6441 we also selected two stars 
(our \# 7 and \#8, right panel of Fig.~\ref{fig2}) 
in common with \citet{gra06} sample (their \#7004050 and \#7004434, respectively) for comparison, 
which lie in a more external field.

The raw two dimensional spectra were processed using the
REDSPEC IDL-based package written
at the UCLA IR Laboratory.
Each order has been
sky subtracted by using the pairs of spectra taken
with the object nodded along the slit, and subsequently
flat-field corrected.
Wavelength calibration has been performed using arc lamps and a second order
polynomial solution, while telluric features have been removed by
dividing by the featureless spectrum of an O star.
At the NIRSPEC resolution of R=25,000 several single
roto-vibrational OH lines and
CO bandheads can be measured to derive accurate oxygen and carbon abundances.
Other metal abundances can be derived from the atomic lines
of Fe~I, Mg~I, Si~I, Ti~I and Ca~I.
Abundance analysis is performed by using full spectral synthesis
techniques and equivalent width measurements of representative lines.

\section{Abundance Analysis}

We compute suitable synthetic spectra
of giant stars by varying the stellar parameters and the
element abundances using an updated
version \citep{ori02} of the code described in \citet{ori93}.
The main features of the code and overall spectral synthesis procedure 
have been widely discussed and tested 
in our previous papers 
and they will not be repeated here.
We only mention that the code 
uses the LTE approximation and is based
on the molecular blanketed model atmospheres of
\citet{jbk80} at temperatures $\le$4000~K
and the ATLAS9 models for temperatures above 4000~K.
The reference solar abundances are from
\citet{gv98}.

Photometric estimates of the stellar parameters are initially used 
as input to produce a grid of model spectra, 
allowing the abundances 
and abundance patterns to vary over a large range and the stellar parameters 
around the photometric values.  
The model which better reproduces the overall observed spectrum and 
the equivalent widths of selected lines is chosen as the best fit model.
We measure equivalent widths in the observed spectrum, in
the best fit model and in four additional models which are, respectively, 
$\pm$0.1 and $\pm$0.2 dex away from the best-fitting, as a further cross-check 
of the inferred abundances and 
overall uncertainties listed in Tables~\ref{tab1},\ref{tab2}.

Stellar parameter uncertainty of $\pm$200~K in temperature (T$_{eff}$), $\pm$0.5
~dex in log-gravity (log~g) and $\pm$0.5~km~s$^{-1}$ 
in microturbulence velocity ($\xi$), 
can introduce a further systematic $\le$0.2~dex uncertainty in the 
absolute abundances.
However, since the CO and OH molecular line profiles are very sensitive to 
effective temperature, gravity, and microturbulence variations, 
they constrain better the values of these parameters,  
significantly reducing their initial range of variation and
ensuring a good self-consistency of the overall spectral
synthesis procedure \citep{ori02,ori04}.

Solutions with  
$\Delta $T$_{\rm eff}$$=\pm$200~K, $\Delta $log~g=$\pm$0.5~dex and
$\Delta \xi$$=\mp$0.5~km~s$^{-1}$ and corresponding 
$\pm$0.2~dex abundance variations from the best-fitting one are indeed less
statistically
significant (typically at $1\le\sigma\le3$ level only, \citet{ori04}). 
Moreover, since the
stellar features under consideration show a similar trend
with variation in the stellar parameters, although with different
sensitivity, {\it relative } abundances are less
dependent on stellar parameter assumptions, 
reducing the systematic uncertainty 
to $<$0.1~dex.
%__________________________________________________________
\begin{table*}
\begin{center}
\caption[]{Photometric and photospheric parameters, heliocentric radial velocity, 
and chemical abundance patterns for the 10 giants observed in 
NGC~6440.}
\label{tab1}
\begin{tabular}{lcccccccccc}
\hline
\hline
star              &\#1 &\#2 &\#3 &\#4 &\#5  &\#6 &\#7 &\#8 &\#9 &\#10\\
 & & & & & & & & & &\\
$\rm (J-K)_0^a$   &1.13 &1.04 &0.86& 1.04&1.03&1.01& 1.07&1.03&1.10&1.08\\
$\rm M_{bol}^a$   &--3.8 &--4.8 &--3.8 &--3.3 &--3.6 &--3.3 &--3.6 &--3.5 &--3.4 &--3.2\\
T$_{\rm eff}$ [K] &3600 &3600 &3800 &3600 &3800 &3800 &3600 &3600 &3600&3600\\ 
log~g             &0.5 &0.5 &0.5 &0.5 &0.5 &0.5 &0.5 &0.5 &0.5 &0.5\\
$v_r$ [km~s$^{-1}$] &--80 &--77 &--86 &--86 &--52 &--62 &--67 &--75 &--71&--77\\
$\rm[Fe/H]$    &--0.53  &--0.59  &--0.65  &--0.50  &--0.55  &--0.53  &--0.55  &--0.41  &--0.56  &--0.52\\
    &$\pm$0.07  &$\pm$0.08  &$\pm$0.10  &$\pm$0.09  &$\pm$0.10  &$\pm$0.10  &$\pm$0.09  &$\pm$0.07  &$\pm$0.09  &$\pm$0.09\\
$\rm[O/Fe]$   & 0.32  & 0.27  & 0.33  & 0.30  & 0.45  & 0.43  & 0.33  & 0.25  & 0.31  & 0.34\\
    &$\pm$0.09  &$\pm$0.11  &$\pm$0.11  &$\pm$0.11  &$\pm$0.17  &$\pm$0.13  &$\pm$0.11  &$\pm$0.09  &$\pm$0.15  &$\pm$0.12\\
$\rm[Ca/Fe]$    & 0.35  & 0.29  & 0.35  & 0.35  & 0.35  & 0.43  & 0.36  & 0.41  & 0.41  & 0.42\\
    &$\pm$0.11  &$\pm$0.12  &$\pm$0.15  &$\pm$0.13  &$\pm$0.13  &$\pm$0.13  &$\pm$0.13  &$\pm$0.11  &$\pm$0.13  &$\pm$0.13\\
$\rm[Si/Fe]$    & 0.23  & 0.29  & 0.25  & 0.34  & 0.35  & 0.43  & 0.35  & 0.31  & 0.36  & 0.32\\
    &$\pm$0.23  &$\pm$0.23  &$\pm$0.19  &$\pm$0.24  &$\pm$0.21  &$\pm$0.21  &$\pm$0.24  &$\pm$0.24  &$\pm$0.24  &$\pm$0.24\\
$\rm[Mg/Fe]$    & 0.31  & 0.28  & 0.30  & 0.33  & 0.35  & 0.40  & 0.35  & 0.33  & 0.34  & 0.33\\
    &$\pm$0.08  &$\pm$0.09  &$\pm$0.10  &$\pm$0.10  &$\pm$0.10  &$\pm$0.10  &$\pm$0.10  &$\pm$0.08  &$\pm$0.10  &$\pm$0.10\\
$\rm[Ti/Fe]$   & 0.23  & 0.19  & 0.25  & 0.30  & 0.40  & 0.43  & 0.40  & 0.26  & 0.46  & 0.42\\
    &$\pm$0.15  &$\pm$0.15  &$\pm$0.19  &$\pm$0.16  &$\pm$0.15  &$\pm$0.15  &$\pm$0.16  &$\pm$0.13  &$\pm$0.16  &$\pm$0.16\\
$\rm[Al/Fe]$   & 0.43  & --    & -- &0.40  & 0.45  & 0.49  & 0.50  & 0.41  & 0.55  & 0.47\\
    &$\pm$0.13  &--        &--        &$\pm$0.14  &$\pm$0.15  &$\pm$0.15  &$\pm$0.14  &$\pm$0.14  &$\pm$0.14  &$\pm$0.14\\
$\rm[C/Fe]$    &-0.37  &-0.41  &-0.25  &-0.30  &-0.25  &-0.27  &-0.45  &-0.29  &-0.54  &-0.48\\
    &$\pm$0.10  &$\pm$0.11  &$\pm$0.12  &$\pm$0.12  &$\pm$0.12  &$\pm$0.12  &$\pm$0.12  &$\pm$0.10  &$\pm$0.12  &$\pm$0.12\\
$\rm^{12}C/^{13}C$    &4.5  &4.0  &6.0  &5.0   &5.0   &5.0   &4.0   &5.0   &3.5   &4.0 \\
    &$\pm$1.r20   &$\pm$1.0   &$\pm$1.6   &$\pm$1.3   &$\pm$1.3   &$\pm$1.3   &$\pm$1.0   &$\pm$1.3   &$\pm$0.9   &$\pm$1.0 \\
\hline
\end{tabular}
\end{center}
$^a$ J,K photometry, E(B-V)=1.15 and (m-M)$_0$=14.58  from \citet{val04}.\\
\end{table*} 
%______________________________________________________________________

%__________________________________________________________
\begin{table*}
\begin{center}
\caption[]{Photometric and photospheric parameters, heliocentric radial velocity, 
and chemical abundance patterns for the 8 giants observed in 
NGC~6441.}
\label{tab2}
\begin{tabular}{lcccccccc}
\hline\hline
star              & \#1 & \#2 & \#3& \#4 &\#5  & \#6 & \#7  & \#8 \\
 & & & & & & & &  \\
$\rm (J-K)_0^a$   &0.85 &0.94 &0.83&0.98&0.99&0.93& 0.87&0.80\\
$\rm M_{bol}^a$   &--4.1 &--3.8 &--3.4 &--3.3 &--3.2 &--3.0 &--2.1 &--2.0\\
T$_{\rm eff}$ [K] &4000 &3800 &4000 &3800 &3800 &3800 &4000 &4000 \\ 
log~g             &1.0 &0.5 &1.0 &1.0 &1.0 &1.0 &1.0 &1.0 \\
$v_r$ [km~s$^{-1}$] &+33 &+12 &-1 &+9 &+13 &+13 &+20 &+9 \\
$\rm [Fe/H]$    &-0.55  &-0.58  &-0.49  &-0.43  &-0.45  &-0.48  &-0.56  &-0.48\\
    &$\pm$0.10  &$\pm$0.08  &$\pm$0.16  &$\pm$0.08  &$\pm$0.08  &$\pm$0.08  &$\pm$0.08  &$\pm$0.08\\
$\rm [O/Fe]$    & 0.28  & 0.35  & 0.26  & 0.23  & 0.17  & 0.23  & 0.29  & 0.33\\
    &$\pm$0.12  &$\pm$0.09  &$\pm$0.17  &$\pm$0.09  &$\pm$0.10  &$\pm$0.10  &$\pm$0.09  &$\pm$0.09\\
$\rm [Ca/Fe]$    & 0.35  & 0.28  & 0.19  & 0.33  & 0.25  & 0.28  & 0.34  & 0.28\\
    &$\pm$0.16  &$\pm$0.13  &$\pm$0.20  &$\pm$0.15  &$\pm$0.15  &$\pm$0.15  &$\pm$0.13  &$\pm$0.13\\
$\rm [Si/Fe]$    & 0.15  & 0.18  & 0.19  & 0.31  & 0.25  & 0.30  & 0.36  & 0.29\\
    &$\pm$0.19  &$\pm$0.20  &$\pm$0.22  &$\pm$0.21  &$\pm$0.21  &$\pm$0.21  &$\pm$0.18  &$\pm$0.18\\
$\rm [Mg/Fe]$    & 0.09  & 0.29  & 0.25  & 0.29  & 0.24  & 0.30  & 0.36  & 0.34\\
    &$\pm$0.11  &$\pm$0.09  &$\pm$0.16  &$\pm$0.09  &$\pm$0.09  &$\pm$0.09  &$\pm$0.08  &$\pm$0.08\\
$\rm [Ti/Fe]$    & 0.35  & 0.12  & 0.29  & 0.33  & 0.35  & 0.38  & 0.36  & 0.28\\
    &$\pm$0.21  &$\pm$0.17  &$\pm$0.24  &$\pm$0.14  &$\pm$0.14  &$\pm$0.14  &$\pm$0.18  &$\pm$0.18\\
$\rm [Al/Fe]$    & 0.60  & 0.51  & 0.51  & 0.53  & 0.55  & 0.58  & 0.43  & 0.49\\
    &$\pm$0.16  &$\pm$0.14  &$\pm$0.20  &$\pm$0.14  &$\pm$0.15  &$\pm$0.14  &$\pm$0.14  &$\pm$0.14\\
$\rm [C/Fe]$    &-0.35  &-0.30  &-0.44  &-0.67  &-0.65  &-0.62  &-0.24  &-0.32\\
    &$\pm$0.13  &$\pm$0.11  &$\pm$0.17  &$\pm$0.11  &$\pm$0.11  &$\pm$0.11  &$\pm$0.11  &$\pm$0.11\\
$\rm^{12}C/^{13}C$    &4.0  &5.0  &5.0  &6.0   &7.0   &6.0   &5.0   &5.0 \\
    &$\pm$1.0   &$\pm$1.3   &$\pm$1.3   &$\pm$1.6   &$\pm$1.8   &$\pm$1.6   &$\pm$1.3   &$\pm$1.3  \\
\hline
\end{tabular}
\end{center}
$^a$ J,K photometry, E(B-V)=0.52 and (m-M)$_0$=15.65  from \citet{val04}.\\
\end{table*} 
%______________________________________________________________________

\section{Results}
\label{results}

By combining full spectral synthesis analysis with equivalent width
measurements, we derive abundances of Fe, C, O
and $^{12}$C/$^{13}$C for the observed
giants in NGC~6440 and NGC~6441.
The abundances of additional $\alpha-$elements, namely Ca, Si, Mg and Ti, are obtained
by measuring a few major atomic lines.
The $^{12}$C/$^{13}$C isotopic ratio has been determined by using the $^{12}$CO and $^{13}$CO 
second overtone bandheads in the H band spectrum covered by our observations
and a few isolated single roto-vibrational lines. 

The near-IR spectra of cool stars also contain many CN molecular lines. 
However, at the NIRSPEC resolution most of these lines are faint and blended 
with the stronger CO, OH and atomic lines, making difficult to impossible a reliable estimate 
of the nitrogen abundance.

Stellar temperatures are both estimated from the $\rm (J-
K)_0$ colors and molecular lines,
gravity from theoretical evolutionary tracks,
according to the location of the stars on the Red Giant Branch (RGB), 
and adopting an average microturbulence velocity of 2.0 km/s
\citep[see also][]{ori97}.
Equivalent widths are computed by Gaussian fitting 
the line profiles, typical values being a few hundreds m\AA,
and the overall uncertainty $\le$10\%.
A table (see Table~\ref{tab3} with the 
measured equivalent widths of representative lines for the observed stars in 
NGC~6440 and NGC~6441 is available in electronic form.

\begin{table*}
\scriptsize
\begin{center}
\caption[]{Log {\it gf} and equivalent widhts  
(m\AA) of some representative lines
for the observed stars in NGC~6440 and NGC~6441.}
\label{tab3}
\begin{tabular}{lccccccccc}
\hline\hline
 & Ca$\lambda $1.61508& Fe~$\lambda $1.61532&Fe~$\lambda $1.55317&Mg~$\lambda $1.57658&Si~$\lambda $1.58884&OH~$\lambda $1.55688&OH~$\lambda $1.55721&Ti~$\lambda $1.55437&Al~$\lambda $1.67634\\
\hline
& \multicolumn{9}{c}{Log {\it gf}}\\
\hline
 &0.362&-0.821&-0.357&0.380&-0.030&-5.454&-5.454&-1.480&-0.55\\ 
\hline
& \multicolumn{9}{c}{Equivalent widths in m\AA}\\
\hline
6440-1  & 220 & 209 & 170 & 395 & 502 & 325 & 331 & 352 & 357 \\
\end{tabular}
\end{center}
\end{table*}

In order to check further the statistical significance of our best-fitting
solution,
we compute synthetic spectra with 
$\Delta $T$_{\rm eff}$$=\pm$200~K, $\Delta $log~g=$\pm$0.5~dex and
$\Delta \xi$$=\mp$0.5~km~s$^{-1}$, and with corresponding simultaneous 
variations 
of $\pm$0.2~dex of the C and O abundances to reproduce the depth of the
molecular features.
We follow the strategy illustrated in \citet{ori04}.
As a figure of merit we adopt
the difference between the model and the observed spectrum (hereafter $\delta$).
In order to quantify systematic discrepancies, this parameter is
more powerful than the classical $\chi ^2$ test, which is instead
equally sensitive to {\em random} and {\em systematic} scatters.
 
Since $\delta$ is expected to follow a Gaussian distribution,
we compute $\overline{\delta}$ and the corresponding standard deviation
for our best-fitting solution and the other models 
with the stellar parameter and abundance variations quoted above.
We then extract 10,000 random subsamples from each
{\it test model} (assuming a Gaussian distribution)
and we compute the probability $P$
that a random realization of the data-points around
a {\it test model} display a $\overline{\delta}$ that is compatible
with an ideal best-fitting model with a $\overline{\delta}$=0. 
$P\simeq 1$ indicates that the model is a good representation of the
observed spectrum.
The statistical tests are performed on portions of the spectra
mainly containing the CO bandheads and the OH lines which are the 
most sensitive to the stellar parameters.

\subsection{NGC~6440}
\label{n6440}

In order to obtain a photometric estimate of the stellar temperatures
and the bolometric magnitudes we use the near IR photometry by \citet{val04} 
and their E(B-V)=1.15 reddening and (m-M)$_0$=14.58 distance modulus. 
We also use the color-temperature transformations and
bolometric corrections of \citet{mon98}, specifically
calibrated for globular cluster giants.
We constrain effective temperatures in the range 3500--4000~K, and
we estimate bolometric magnitudes 
$\rm M_{bol}\le -3.2$ 
(see Table~\ref{tab1}).
The final adopted temperatures, obtained by best-fitting the CO and in 
particular the OH molecular bands which are 
especially temperature sensitive in cool giants, are also reported in
Table~\ref{tab1}.

%________________________________________________________________
\begin{figure}
\centering
\includegraphics[width=9cm]{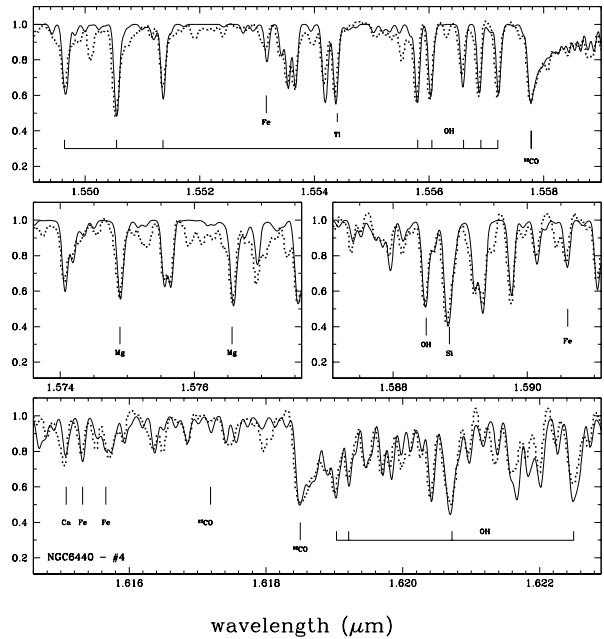}    
\caption{
Selected portions of the observed echelle spectra (dotted lines) of star \#4
in NGC~6440 with our best-fitting synthetic spectrum
(solid line) superimposed. A few important molecular and atomic lines
are marked.
}
\label{fig3}
\end{figure}
%________________________________________________________________

Fig.~\ref{fig3} shows
our synthetic best fit superimposed on the
observed spectra of giant \#4 in NGC~6440.
From our overall spectral analysis 
we find average [Fe/H]$=-0.50\pm0.02$, [O/Fe]$=0.33\pm 0.02$ and 
$\rm [\alpha/Fe]=0.34\pm0.02$
(see Table~\ref{tab1}).
We also measure an average carbon depletion ([C/Fe]=--0.36$\pm0.03$~dex)
and low $\rm ^{12}C/^{13}C\approx 4.6\pm 0.7$ isotopic ratio. 

Our best-fitting solutions  
have an average probability $\-P>$0.99 to be statistically 
representative of the observed spectra. 
The other {\it test models} with different assumptions for the stellar 
parameters are only significant at $\approx$1$\sigma$ 
(warmer stars with corresponding 0.2~dex higher abundances) and 3$\sigma$ 
(cooler stars with corresponding 0.2~dex lower abundances) level. 

From the NIRSPEC spectra we also derived stellar heliocentric radial velocities
(see Table~\ref{tab1}), finding an average value 
$\rm v_r=-74\pm $4~km/s with a dispersion $\sigma=11\pm3$~km/s.
By excluding star \#5, which is the most discrepant, we find   
$\rm v_r=-76\pm $3~km/s and $\sigma=9\pm2$~km/s.
These values are in good agreement with the one quoted in \citet{har96}.

\subsection{NGC~6441}
\label{n6441}

We use the near IR photometry of \citet{val04}  
and their E(B-V)=0.52 reddening and (m-M)$_0$=15.65 distance modulus. 
We find photometric temperatures in the 3800-4200~K range and bolometric 
magnitudes between -2.0 and -4.1 (see Table~\ref{tab2}).
The final adopted temperatures, 
obtained by best-fitting the CO and the OH molecular bands, 
are also reported in Table~\ref{tab2}.
%________________________________________________________________
\begin{figure}
\centering
\includegraphics[width=9cm]{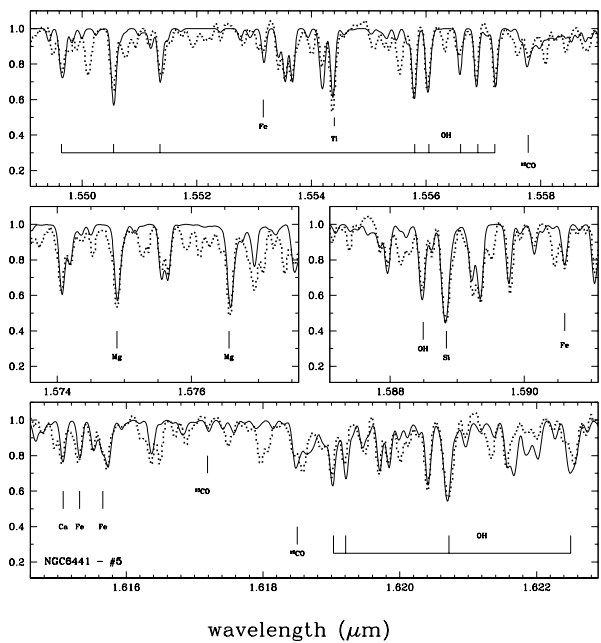}    
\caption{
Selected portions of the observed echelle spectra (dotted lines) of star \#5 
in NGC~6441 with our best-fitting synthetic spectrum
(solid line) superimposed. A few important molecular and atomic lines
are marked.
}
\label{fig4}
\end{figure}
%________________________________________________________________

Fig.~\ref{fig4} shows
our synthetic best-fitting superimposed on the
observed spectra of giant \#5 in NGC~6441.
For this cluster our abundance analysis gives an average [Fe/H]$=-0.50\pm0.02$,
[O/Fe]$=+0.27\pm0.02$ and an overall average $\rm [\alpha/Fe]=0.28\pm0.02$.
We also measure an average carbon depletion ([C/Fe]=--0.45$\pm0.06$~dex)
and a low $^{12}$C/$^{13}$C$\approx5.4\pm0.9$. 

Our iron abundance for stars \#7 and \#8 is only 0.10~dex lower and 
our average abundance
0.11~dex lower than in \citet{gra06}, so fully consistent 
within the errors.
Our estimates for the other abundances are also
similar (within $\pm$0.1 dex) with those quoted by \citet{gra06}. 
Only Ca, which is rather low in \citet{gra06} compared to the other $\alpha$-elements,
is somewhat more discrepant ($\approx$0.2 dex) but still barely consistent 
within the errors. 

In order to further check the robustness of our best-fitting solutions,  
the same statistical test done for NGC~6440 has been repeated here.
Our best-fitting solutions  
have an average probability $P>$0.99 to be statistically representative 
of the observed spectra, while the other {\it test models} are only 
significant at $\ge 1~\sigma$ level. 

By measuring the stellar radial velocities (see Table~\ref{tab2})
we find an average $\rm v_r=+13\pm$3~km/s with a dispersion $\sigma=10\pm2$~km/s, 
somewhat lower, but still in reasonable agreement within the errors, with the values proposed by \citet{gra07} 
based on a sample of stars three times larger, namely 
$\rm v_r$=+21~km/s and $\sigma$=13~km/s.
Our radial velocity is in excellent agreement with the $\rm v_r$=+15~km/s estimated 
by \citet{dub97}.

%________________________________________________________________
\begin{figure}
\centering
\includegraphics[width=9cm]{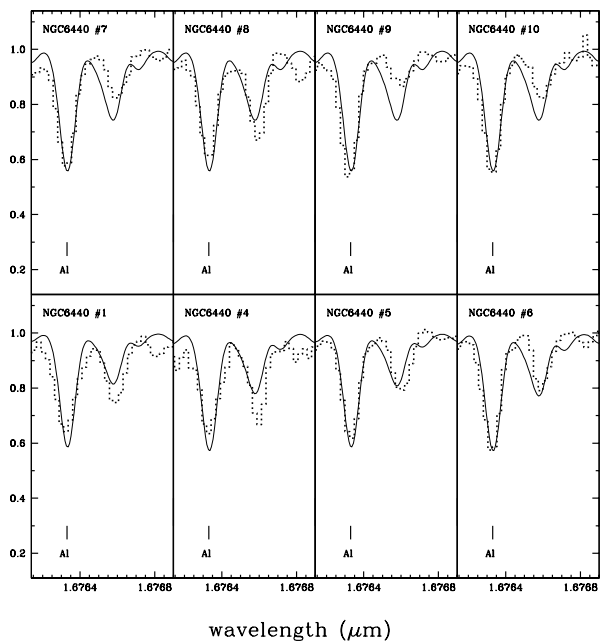}    
\caption{
Observed spectra in the Al line region at 1\.67 $\mu$m, of the stars in  
NGC~6440 with our best-fitting synthetic spectrum
(solid line) superimposed. 
}
\label{fig5}
\end{figure}
%________________________________________________________________

%________________________________________________________________
\begin{figure}
\centering
\includegraphics[width=9cm]{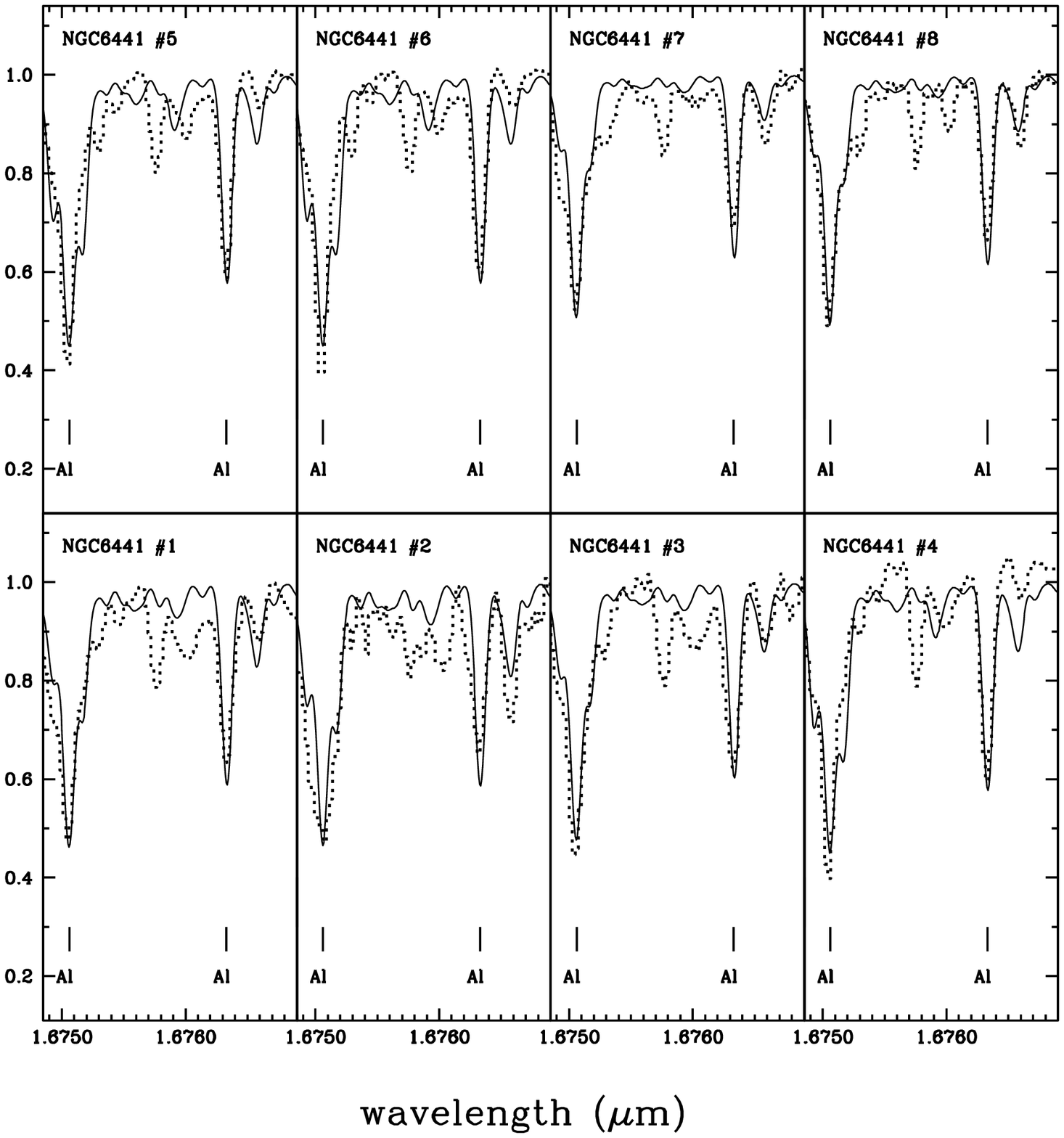}    
\caption{
Observed spectra in the Al line region at 1\.67 $\mu$m, of the stars in  
NGC~6441 with our best-fitting synthetic spectrum
(solid line) superimposed. 
}
\label{fig6}
\end{figure}
%________________________________________________________________

\subsection{Aluminum lines}

%________________________________________________________________
\begin{figure}
\centering
\includegraphics[width=9cm]{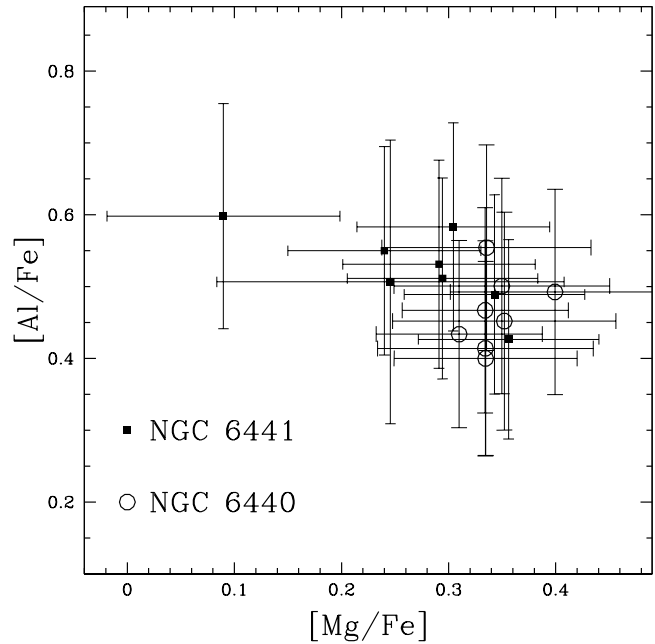}    
\caption{
[Mg/Fe] {\it vs} [Al/Fe] abundance ratio distribution for the observed stars 
in NGC~6440 (open circles) and   
NGC~6441 (filled squares).
}
\label{fig7}
\end{figure}
%________________________________________________________________

The Al doublet at $\rm \lambda\lambda1.67506,1.67634$ is at one side of the NIRSPEC-5 echellogram and 
easily measurable measurable in stars with positive radial velocities as for the giants in NGC~6441, 
while only the reddest line is barely measurable in the giants of NGC~6440.
These lines have been proved to be reliable Al abundance indicators, having 
relatively high excitation potential and
forming quite deeply in the stellar atmospheres, where the LTE regime
still dominates, even in giants with low temperatures and gravities
\citep[see e.g.][]{bg96}.
When available, the two lines of the doublet provide very similar Al abundances within the errors. 
However, since the $\lambda1.67506$ line is very close to the edge of the echellogram, we 
only use it as a double check of the Al  abundance obtained from the $\lambda1.67634$.

Figs.\ref{fig5},\ref{fig6} show the observed spectra in the Al region with overimposed our best fit.
The spectra of stars \#2 and \#3 in NGC~6440 were too noisy at 1.67 $\mu$m to be used for reliable Al 
abundance determinations.

We find 
average $\rm [Al/Fe]=0.45\pm0.02$ and
$\rm [Al/Fe]=0.52\pm0.02$ in NGC~6440 and NGC~6441, respectively 
(see also Tables \ref{tab1} and \ref{tab2}). 
Fig.\ref{fig7} shows the plot of the [Mg/Fe] {\it vs} [Al/Fe] distribution 
for the two clusters.
Although our samples are too small to draw firm conclusions, we find 
some evidence of an anti-correlation between the two abundance ratios in NGC~6441, 
with a Spearman rank correlation coefficient of -0.7, corresponding to a probability 
of only $\approx$5\% that the two variables be not correlated. 
Such an evidence is not presently detected in NGC~6440.

\section{Discussion and Conclusions}

Our high resolution spectroscopy in the near IR gives 
radial velocities well in agreement with previous studies, 
and typical ($|v_r|\le \pm 100$ km/s) of the bulge population. 
Also the inferred velocity dispersions are fully consistent 
with previous estimates and 
typical of massive globular clusters. 
Our iron abundances for 
NGC~6440 and NGC~6441  
confirm the 1/3 solar abundance derived from photometric
estimates, as obtained from the RGB morphology and luminosity in the optical 
as well as in the near IR. 

The enhancement of the [$\alpha$/Fe] abundance ratio, relative to the solar value,
agrees well with other studies of clusters and bulge field. 
This reinforces the statement of a ab-origin enrichment of the interstellar medium 
by type SNII and 
an overall star formation process and chemical enrichment on a 
relatively short timescale.  

The low $\rm^{12}C/^{13}C$ isotopic abundance ratios measured in NGC~6440 and NGC~6441
are similar to those measured in other giant stars of the 
bulge \citep{ori02,she03,ori04} and the other globular clusters  
of our survey, as well as in the halo clusters
\citep{she96,gra00,vws02,smi02,ori03},
suggesting that additional mixing mechanisms due to 
{\it cool bottom processing} in the stellar interiors during the evolution along
the RGB \citep[see e.g.][]{cha95,dw96,csb98,bs99} occurs also at high metallicity.

O-Na and Mg-Al anti-correlations have been recently discovered in several massive globular
cluster stars \citep[see e.g.][ and references therein]{sne04,john05,gra04,car06}.
Recently, \citet{gra06,gra07} also find evidence of such anti-correlations in NGC~6441.
Such star-to-star abundance variations of O, Na, Al and Mg are currently interpreted as 
inhomogeneities in the proto-cluster gas, due to the pollution of a former generation 
of massive AGB stars.
Indeed, the H-burning at high temperatures, 
as occurring in thermally pulsing AGB undergoing
hot bottom burning \citep{ven01,gra04}, allows the O-N, Ne-Na
and Mg-Al nucleosynthesis cycles\citep{lan93}, 
although the latter requires higher temperatures than the previous ones, 
and produces less prominent abundance variations.
Na-Al-rich and O-Mg poor globular cluster stars
are thus expected to have formed within the kinematically
cool ejecta of such a first generation of massive AGB stars 
\citep[e.g.][]{cot81,coh02}.

\section*{Acknowledgments}

LO acknowledge the financial support by 
the Ministero dell'Istru\-zio\-ne, Universit\`a e Ricerca (MIUR).

RMR acknowledges support from grant number AST-0709479,
from the National Science Foundation.
The authors are grateful to the staff
at the Keck Observatory and to Ian McLean
and the NIRSPEC team.
The authors wish to recognize and acknowledge the very significant cultural
role and reverence that the summit of Mauna Kea has always had within
the indigenous Hawaiian community.
We are most fortunate to have the opportunity to conduct observations 
from this mountain.

The authors acknowledge the anonymous Referee for his/her useful comments.

\label{lastpage}
\end{document}